# Topotactic phase transformation in correlated vanadium dioxide through oxygen vacancy ordering

*Xuanchi Zhou [1, 2] \*, Xiaohui Yao [1], Xiaomei Qiao [1], Jiahui Ji [1], Guowei Zhou [1, 2] \*, Huihui Ji [1, 2], Xiaohong Xu [1, 2] \**

[1] *Key Laboratory of Magnetic Molecules and Magnetic Information Materials of Ministry of Education & School of Chemistry and Materials Science, Shanxi Normal University, Taiyuan, 030031, China*
[2] *Research Institute of Materials Science, Shanxi Key Laboratory of Advanced Magnetic Materials and Devices, Shanxi Normal University, Taiyuan 030031, China*

\*Authors to whom correspondence should be addressed: *xuanchizhou@sxnu.edu.cn (X. Zhou)*, *zhougw@sxnu.edu.cn (G. Zhou)*, and *xuxh@sxnu.edu.cn (X. Xu)*.

**Abstract**

Controlling the insulator-metal transition (IMT) in correlated oxide system through oxygen vacancy ordering opens up a new paradigm for exploring exotic structural transformation and physical functionality. Oxygen vacancy serves as a powerful tuning knob for adjusting the IMT property in $VO_2$, though driving topochemical reduction to $V_2O_3$ remains challenging due to structural incompatibility and competing phase instability. Here we unveil consecutive oxygen-vacancy-driven $VO_2$-$VO_{2-x}$-$V_2O_3$ topotactic phase transformation route with enticing facet-dependent anisotropy, engendering tunable IMT properties over an extended temperature range. Remarkably, topochemically reduced $V_2O_3$ inherits the crystallographic characteristics from parent $VO_2$, enabling emergent lattice framework and IMT behavior inaccessible via direct epitaxial growth. Analogous electron doping arising from hydrogenation and oxygen vacancy contributes cooperatively to drive the Mott phase transition in $VO_2$ through band-filling control. Our work not only unveils sequential topotactic phase transformations in $VO_2$ through oxygen vacancy ordering but also provides fundamentally new insights for defect-mediated Mott transitions.

**TOC Figure**:

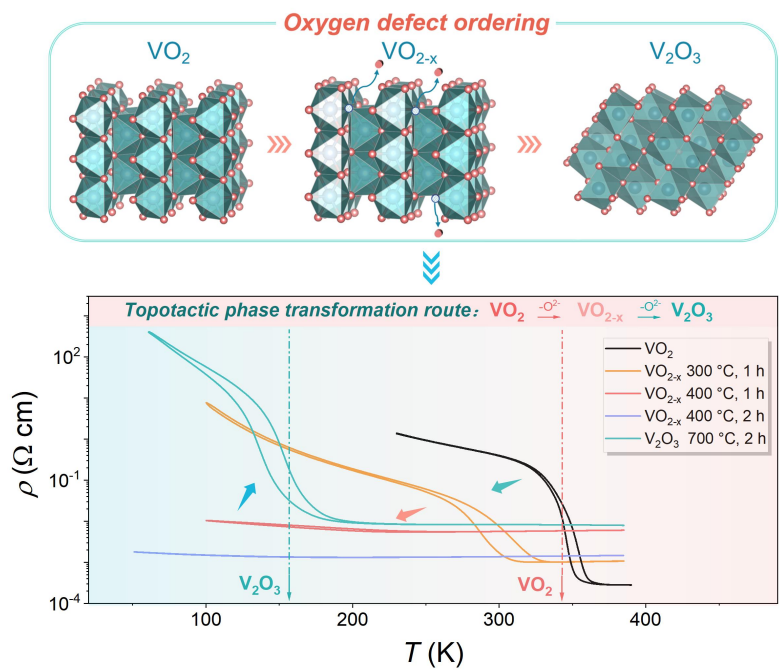

Oxygen defects, functioning as pivotal control parameters in correlated oxide system, initiated an emerging chapter to drive topotactic phase transformations with emergent physical functionality and phenomena, which is harnessed in energy conversion and storage,[1] superconductivity,[2,3] ferroelectric,[4,5] and correlated electronics.[6] Controlling the oxygen defects in correlated oxides offers a rewarded pathway to exploit the cation displacement and oxygen octahedron rotation or tilt for modifying the orbital overlapping and bandwidth.[7] Accompanied by structural evolution, the incorporation of oxygen vacancy intrinsically donates two electrons into the conduction band of correlated oxides, which tailors the band filling and electronic orbital configuration.[8] Benefiting from the ion-lattice-charge-spin coupling, the spin ordering and magnetic exchange interactions in oxygen-deficient correlated oxides can be manipulated through local symmetry breaking, discovering novel magnetoelectric ground states.[9,10] Notable examples include perovskite $SrCoO_3$,[9] $SrFeO_3$ [11] and $CaFeO_3$,[12] in which oxygen vacancy ordering stabilizes emergent brownmillerite and infinite-layer structures hosting distinct magnetoelectric properties via thermodynamic stability modulation, exemplifying the irreplaceable role of oxygen defect in regulating physical property. Typically, oxygen ionic evolution triggers topotactic phase transformation in correlated $SrCoO_3$ system from ferromagnetic metal ground state to antiferromagnetic insulator $SrCoO_{2.5}$,[13] while the unconventional superconductivity was discovered in infinite-layer $Nd_{0.8}Sr_{0.2}NiO_2$ via $CaH_2$ topochemical reduction.[2] Therefore, oxygen vacancy as an indispensable ingredient provides a powerful tool for exploring exotic physical functionalities and phenomena in correlated electron system.

Unlike corner-sharing $BO_6$ octahedra in perovskites, binary oxides with edge- or face-sharing octahedral frameworks offer a fertile ground to explore new physical functionality via oxygen vacancy ordering, due to direct cation-cation interaction and structural flexibility. As a representative case, vanadium dioxide ($VO_2$) undergoes a temperature-controlled first-order insulator-metal transition (IMT) governed by the critically balanced Peierls-Mott physics that is readily tunable by oxygen stoichiometry, providing a quintessential testbed for defect engineering.[14-16] Conventionally, oxygen vacancy was unveiled to enlarge the relative stability in metallic orbital configuration of $VO_2$ [17,18] or trigger the complete metallization [19] due to the donation of electrons to the conduction band, during which the lattice framework remains unchanged. It is in particular worthy to note that the phase diagram related to vanadium oxides is exceptionally complex, rooted in the multivalence nature of vanadium, with diverse vanadium oxides displaying distinct IMT behaviors spanning a wide temperature range.[20,21] One focal challenge persists whether oxygen-vacancy-mediated chemical reduction of $VO_2$ can drive topotactic phase transformation to ordered $V_2O_3$, structurally akin to perovskite-to-brownmillerite evolution, given inherent barriers from the competing phase instabilities and structural mismatch. In addition, the ionic interaction between proton and oxygen vacancy has sparked a surge of research interest, in which oxygen ionic migration is kinetically expedited by hydrogenation.[22,23] Beyond that, hydrogenation introduces analogous electron doping to $VO_2$, raising a fundamental question: Can protonic-mediated and oxygen-vacancy-driven electron doping cooperatively drive Mott phase transition through band filling?

In this letter, we showcase the precise control over the defect engineering in correlated $VO_2$ system, in which the IMT property of $VO_2$ is progressively depressed, followed by a topotactic phase modulation to $V_2O_3$ through oxygen vacancy ordering. Remarkably, such the topochemical reduction from $VO_2$ to $V_2O_3$ exhibits facet-dependent anisotropy while inheriting the crystallographic characteristics from parent $VO_2$. In addition, cooperative electron doping effect is demonstrated in correlated $VO_2$ system through hydrogenation and oxygen defects, jointly regulating the IMT property via coupled band-filling control. Our findings not only experimentally observe sequential $VO_2$-$VO_{2-x}$-$V_2O_3$ topotactic phase modulation through oxygen vacancy ordering, with facet-dependent anisotropy, but also clarify the underneath mechanism governing filling-controlled Mott phase transitions in correlated electron system.

Controlling the chemical reduction through defect engineering offers a powerful tool for manipulating the structural transformation and physical functionality in correlated electron system. Typically, oxygen-vacancy-driven topotactic phase transformation from perovskite to brownmillerite and even infinite-layer structure results in enticing magnetoelectric coupling in perovskite oxide system with ordered oxygen vacancy channel, bringing in emergent physical functionality (Figure 1a).[9] Beyond structural evolution, oxygen vacancy directly donates electrons to the conduction band of $VO_2$ to elevate the Fermi level ($E_F$) and stabilize the metallic orbital configuration, thereby reducing the transition temperature ($T_{IMT}$) or triggering collective carrier delocalization through band-filling control. Unlike topotactic perovskite-to-brownmillerite reduction enabled by planar oxygen-vacancy ordering, achieving sequential two-step $VO_2$-$VO_{2-x}$-$V_2O_3$ transition remains fundamentally challenged by the competing phase instabilities and rutile-corundum structural mismatch (Supplementary Note 1). While electron-beam illumination enables nanoscale chemical reduction of $VO_2$ to $V_2O_3$,[24] this spatially confined approach limits the scalability and precludes uniform topotactic phase transformation across macroscopic film, rendering the absence of sharp IMT behavior. Beyond that, filling-controlled Mottronics offers an alternative pathway to engineer the physical functionality of correlated $VO_2$, where joint electron doping arising from the oxygen and hydrogen ionic evolution is poised to synergistically drive electronic phase transitions.

To precisely introduce the oxygen deficiency in correlated $VO_2$ system, a high-vacuum annealing strategy (e.g., $P_{O_2} \sim 1 \times 10^{-5}$ Pa) is herein exploited, driving the oxygen desorption through thermodynamic equilibration (Supplementary Note 2). The introduction of oxygen vacancy renders an approximately 3 % *out-of-plane* lattice expansion in VO$_2$, evidenced by a leftward shift related to the (002) plane diffraction peak from 40.11 º to 38.88 º, the magnitude of which is similar to previous study (Figure 1b).[17, 18] It is worthy to point out that the small diffraction shoulder detected at 38.05 º likely signifies the emergence of metastable Magnéli phase in oxygen-deficient VO$_{2-x}$ during the high-vacuum annealing, consistent with previous report.[25] Further elevating the high-vacuum annealing temperature to 700 ºC instead results in the topochemical reduction of VO$_2$ toward V$_2$O$_3$, in which the characteristic diffraction peaks associated with the (110) and (006) planes of V$_2$O$_3$ are clearly identified. This observation starkly differs from the direct V$_2$O$_3$ epitaxial growth on *c*-plane sapphire, in which case the grown V$_2$O$_3$ film displays exclusive *out-of-plane* preferential orientation enabled by the corundum-to-corundum coherent epitaxy. Critically, topochemical reduction permits the inheritance of crystallographic orientations from parent VO$_2$ grown on *c*-plane Al$_2$O$_3$, where the symmetry mismatch between hexagonal Al$_2$O$_3$ and rutile VO$_2$ induces the domain matching epitaxy, yielding equivalent twin variants in VO$_2$ film.[26] As a result, the (110)-oriented V$_2$O$_3$ films emerge via topochemically reducing $(01\bar{1})_R$-oriented domains in parent VO$_2$, crystallographically equivalent to the (110) orientation. Further consistency in the crystalline structure of topochemically reduced V$_2$O$_3$ is further verified by using high-resolution transmission electron microscopy (HRTEM), where coherent heterointerface epitaxy is validated (Figure 1c), with a corundum structure being detected through Fast Fourier transform (FFT) (Figure 1d). The film thickness of topochemically reduced V$_2$O$_3$ is identified as around 25.40 nm using HRTEM technique (Figure S1). Topochemically reduced V$_2$O$_3$ retains distinct twin variants with (001) and (110) orientations, a direct evidence of structural inheritance from its parent VO$_2$, in accord with XRD result (Figures 1e-1f).

Sequential topotactic phase modulations in correlated VO$_2$ system via oxygen vacancy ordering are further confirmed by respective temperature dependence of material resistivity ($\rho$-*T* tendency) in Figure 2a. It is found that pristine VO$_2$ undergoes sharp IMT behavior at the critical temperature of ~339 K, while the introduction of oxygen vacancy progressively suppresses the IMT property of VO$_2$, reducing the $T_{IMT}$ toward room temperature. Performing the high-vacuum annealing at 400 ºC for 1 h instead renders the collective metallization of VO$_2$ while prolonging the annealing period to 2 h further reduces the material resistivity. In striking contrast, elevating the annealing temperature to 700 ºC drives the topochemical reduction from VO$_2$ to V$_2$O$_3$, which exhibits a thermally-driven IMT behavior at 132 K, as determined by respective temperature coefficient of resistance (*TCR*)-*T* tendency (Figure S2). Across the IMT, topochemically reduced V$_2$O$_3$ exhibits a resistivity modulation exceeding 5 orders of magnitude, significantly surpassing the minor resistivity upturn observed in electron-beam-illuminated V$_2$O$_3$,[24] with a $T_{IMT}$ (132 K) below the bulk value (~156 K). Notably, this observation strongly differs from the extensively depressed IMT in epitaxially grown V$_2$O$_3$ on *c*-facet sapphire,[27] due to the weakened self-induced strain from pre-existing (110)-oriented domains. The above findings clearly demonstrate two-step VO$_2$-VO$_{2-x}$-V$_2$O$_3$ topotactic phase transformation within $VO_2$ system through oxygen vacancy ordering, incurring a variety of IMT behaviors across a wide temperature range.

Moreover, topochemically reduced $V_2O_3$ films retain robust chemical stability under ambient conditions, as demonstrated by the temporal stability of characteristic diffraction peak and material resistance (Figure 2b). It is well-known that the Raman peaks at 194 cm$^{-1}$ ($\omega 1$) and 223 cm$^{-1}$ ($\omega 2$) represent the V-V dimer vibrations, while the Raman peak located at 612 cm$^{-1}$ ($\omega 3$) arises from V-O bond length asymmetry (Figure 2c).[28] The above characteristic Raman peaks, structural fingerprints of insulating $VO_2$, are depressed for oxygen-deficient $VO_{2-x}$, aligning well with the metallization of $VO_2$ through oxygen vacancy. By contrast, topochemically reduced $V_2O_3$ exhibits emergent Raman peaks corresponding to the $E_g$ and $A_{1g}$ vibrational modes of corundum $V_2O_3$, unveiling that parent $VO_2$ transforms to a reduced $V_2O_3$ phase. The above findings demonstrate oxygen-vacancy-driven two-step topotactic phase transformation route associated with $VO_2$-$VO_{2-x}$-$V_2O_3$, transcending conventional vacancy-mediated metallization in $VO_2$. To probe the changes in chemical environment of $VO_2$ through oxygen vacancy, X-ray photoelectron spectra (XPS) analysis was performed, as the V $2p$ and O $1s$ core-level spectrum shown in Figure 2d-2e, respectively. The valence state of vanadium is extensively reduced from $V^{4+}$ toward $V^{3+}$ via defect engineering, particularly for topochemically reduced $V_2O_3$. The introduction of oxygen vacancy for $VO_2$ is evidenced by the elevated intensity of defect-associated oxygen species relative to the lattice oxygen, while subsequent structural ordering into the corundum $V_2O_3$ mitigates the pre-existing oxygen defects.

Notably, the facet-dependent anisotropy in topochemical reduction via oxygen defect is demonstrated in $VO_2$ system, wherein topotactic phase transformation from $VO_2$ to $V_2O_3$ fails to initiate on $a$-facet $TiO_2$ template, in contrast with the $c$-facet and $r$-facet sapphire templates (Figure 3a). Unlike direct epitaxy growth, topochemically reduced $V_2O_3$ preserves the crystallographic characteristics from parent $VO_2$,[29] for example, yielding a $V_2O_3$ film on $c$-facet sapphire with distinct (110)- and (001)-oriented twin domains stabilized through domain-matching epitaxy in $VO_2/Al_2O_3$ (0001) precursor (Figure S3).[30] Furthermore, consecutive topotactic phase transformations are analogously realized in the $r$-facet sapphire template, while topochemically reduced $V_2O_3$ grown on $r$-facet $Al_2O_3$ substrate adopts a (012) preferential orientation, consistent with the direct epitaxy growth of $V_2O_3$ (Figure S4). By contrast, the grown $VO_2$ on $a$-facet $TiO_2$ template cannot transforms into a reduced $V_2O_3$ phase under identical high-vacuum annealing, exhibiting only a left-shifted diffraction peak of $VO_2$ that overlaps the substrate signature. This finding is in consistency with corresponding electrical transport properties, wherein the characteristic IMT behavior of $V_2O_3$ is observed for topochemically reduced films deposited on the $r$- and $c$-facet $Al_2O_3$ substrates (Figure 3b). Nevertheless, for films grown on the $a$-facet $TiO_2$, the electrical transport behaviors still resemble that of $VO_2$, with a lowered $T_{IMT}$ being observed, indicating the presence of oxygen-deficient $VO_{2-x}$ rather than the conversion to $V_2O_3$. The anisotropy in topochemical reduction of $VO_2$ stems from an enlarged vacancy formation energy on *a*-facet $TiO_2$ substrates, which kinetically suppresses the formation of abundant oxygen vacancy.[24] Analogous variations in the valence states for $VO_2$ grown on *r*-facet sapphire via topochemical reduction are characterized using XPS analysis (Figure S5), while respective surface topology is identified by the atomic force microscope (Figure S6). Furthermore, the $T_{IMT}$ achievable in topochemically reduced $V_2O_3$ on *c*-facet $Al_2O_3$ substrate is lower than the one deposited on *r*-facet $Al_2O_3$ due to different strain states imposed by epitaxial templates, in agreement with previously reported epitaxial growth of $V_2O_3$ (Figure 3c).[31] Topochemical reduction of $VO_2$ to $V_2O_3$ through oxygen vacancy ordering unlocks novel crystallographic orientations and IMT property unattainable through conventionally direct epitaxial deposition, with high reproducibility (Figure S7).

Over the past decade, filling-controlled Mott phase modulation has emerged as a groundbreaking paradigm for exploring exotic electronic states and physical functionality in correlated $VO_2$ system, covering hydrogenation and oxygen vacancy, beyond thermally-driven phase transition (Figure 4a). A fundamental question thus emerges: Can co-engineered hydrogenation and oxygen vacancy, acting as electron donors, collaboratively drive the Mott phase modulations through coupled band filling. To address this hypothesis, oxygen-deficient $VO_{2-x}$ sample (e.g., 300 ºC, 1 h) was further hydrogenated through hydrogen spillover strategy, in which the sputtered Pt catalyst significantly reduces the energy barrier for dissociating $H_2$ molecules into protons and electrons.[32, 33] Oxygen-deficient $VO_{2-x}$ still retains a suppressed IMT property, manifested by a markedly reduced $T_{IMT}$ approaching 285.5 K and depressed transition sharpness. Nevertheless, further performing a mild hydrogenation at 70 ºC for 30 min engenders the complete metallization of $VO_{2-x}$, unveiling joint electron doping effect from hydrogenation and oxygen vacancy in driving Mott phase transition (Figure 4b). In addition, such the cooperative electron doping effect is in agreement with respective structural evolution in Figure S8, wherein the diffraction peak associated with (020) plane of oxygen-deficient $VO_{2-x}$ progressively shifts leftwards with subsequent hydrogenation.

To probe the changes in electronic orbital configuration of $VO_2$ via ionic evolution, soft X-ray absorption spectroscopy (sXAS) technique was performed, as their V-*L* edge and O-*K* edge shown in Figures 4c-4d, respectively. The V-*L* edge spectrum associated with the V $2p \rightarrow 3d$ transition is widely recognized as a sensitive indicator of the vanadium valence state.[34] The introduction of oxygen vacancy induces a lower-energy shift in both V-$L_{III}$ and V-$L_{II}$ peaks for oxygen-deficient $VO_{2-x}$, an effect amplified by subsequent hydrogenation (Figure 4c). The O *K*-edge spectrum can provide an indirect probe of the density of states in unoccupied conduction band for $VO_2$, where the relative variation in the first (second) peak intensity tracks the electron occupation in the $t_{2g}$ ($e_g$) band, given the empty O-$2p$ states and the V-$3d$ and O-$2p$ orbital hybridization.[35] Oxygen vacancy extensively reduces the relative peak intensity of first peak derived from the $t_{2g}$ orbital, indicating the electron occupation in the low-energy $t_{2g}$ orbital, with hydrogenation further intensifying this effect (Figure 4d). These findings demonstrate joint electron doping mediated by hydrogenation and oxygen vacancy in driving orbital reconfiguration towards electron-itinerant state through synergistic band filling, a phenomenon also validated in W-doped $VO_2$ material (Figures S9-S12). It is worthy to point out that such the coupled electron-doping effect in driving Mott phase modulation is non-trivial, for example, chemical doping tends to suppress the electron localization in hydrogenated $VO_2$ despite similar carrier doping.[36]

Beyond that, the intercalated protons preferentially interact with engineered oxygen vacancies, while such the ionic interactions are poised to effectively modulate the oxygen ionic migration.[22] With the oxidative annealing in air at 200 ºC for 1 h, oxygen-deficient $VO_{2-x}$ gradually reverts to its initial electronic state due to the extraction of oxygen defects, accompanied by an expected IMT behavior (Figure 4f), consistent with respective structural evolution (Figure S13). Notably, the recovery in the transport property under oxidative annealing is more pronounced for oxygen-deficient $VO_{2-x}$ through hydrogenation, in comparison with non-hydrogenated counterpart (Figure 4g). This trend aligns with the structural transformation characterized by using Raman spectroscopy, in which situation hydrogenated $VO_{2-x}$ under oxidative annealing exhibits more intensified Raman signatures associated with the V-V dimers of insulating $VO_2$ (Figure 4h). The electrons introduced by hydrogenation tend to accumulate around the oxygen vacancy within the lattice of $VO_2$, thereby expediting the oxygen migration kinetics. The above findings not only unveil the overlooked cooperative electron-doping effect resulting from hydrogenation and defect engineering in driving Mott phase transition within correlated electron system but also unveil an expedited oxygen ionic transport through ionic interactions.

In this letter, we demonstrate sequential topochemical reduction in correlated $VO_2$ system through oxygen vacancy ordering, driving topotactic phase transformation along the $VO_2$-$VO_{2-x}$-$V_2O_3$ pathway. Oxygen vacancy initially enlarges the relative stability in the metallic orbital configuration of $VO_2$ to reduce the $T_{IMT}$ or trigger collective carrier delocalization, followed by a topotactic phase transformation into $V_2O_3$ with enticing facet-dependent anisotropy. Topochemical reduction from $VO_2$ to $V_2O_3$ not only extensively broadens the design landscape for binary oxides harnessing oxygen vacancy ordering but also retains crystallographic characteristics from parent $VO_2$ to unlock emergent functionality unattainable via direct epitaxial growth. Critically, cooperative electron donation arising from defect engineering and hydrogenation is unveiled to jointly drive Mott phase modulation in $VO_2$ via band-filling control, while respective ionic interactions kinetically accelerate the oxygen ionic transport of $VO_2$. Our work establishes a promising pathway to harness functionalities associated with topochemical reduction in correlated electron system through oxygen vacancy ordering, while delivering fundamental insights into defect-mediated Mott transitions.

**Notes**

The authors declare no competing financial interest.

**Acknowledgements**

This work was supported by the National Natural Science Foundation of China (Nos. 52401240, U24A6002, 52471203, and 12174237), Fundamental Research Program of Shanxi Province (No. 202403021212123), Scientific and Technologial Innovation Programs of Higher Education Institutions in Shanxi (No. 2024L145), Shanxi Province Science and Technology Cooperation and Exchange Special Project (No. 202404041101030) and the Open Project of Tianjin Key Laboratory of Optoelectronic Detection Technology and System (No. 2024LODTS102). The authors also acknowledge the beam line BL08U1A at the Shanghai Synchrotron Radiation Facility (SSRF) (https://cstr.cn/31124.02.SSRF.BL08U1A) and the beam line BL12B-b at the National Synchrotron Radiation Laboratory (NSRL) (https://cstr.cn/31131.02.HLS.XMCD.b) for the assistance of sXAS measurement.

**Figures and captions**

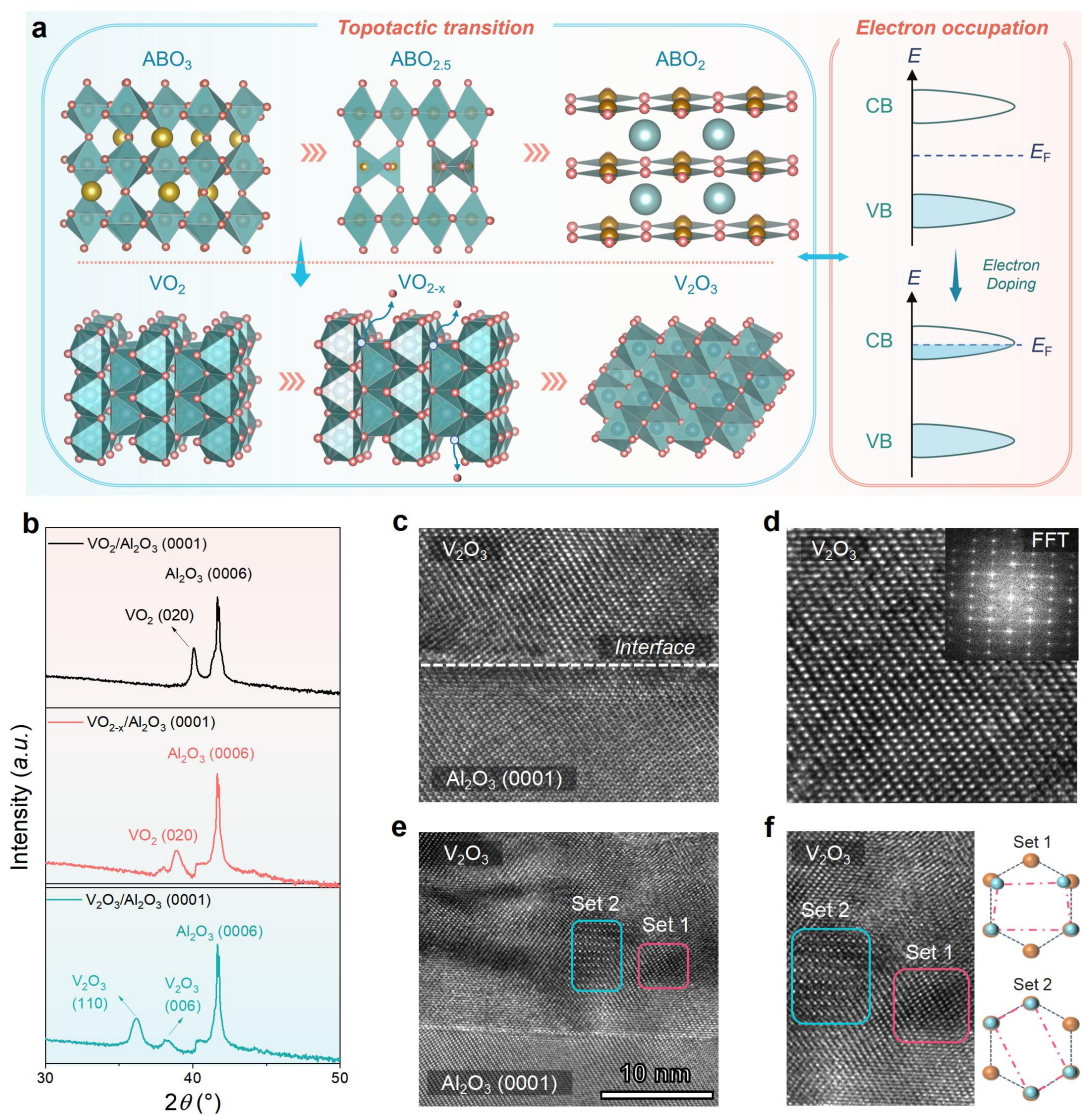

**Figure 1. Topotactic structural transformation in VO$_2$ through oxygen vacancy. a**, Schematic of topotactic phase transformation and band filling control in perovskite oxides and binary oxides. In addition, Conduction band (CB) and valence band (VB) are denoted accordingly. The ABO$_3$, ABO$_{2.5}$ and ABO$_2$ denote the perovskite, brownmillerite and infinite-layer structures, respectively. **b**, X-ray diffraction (XRD) patterns compared for pristine VO$_2$, oxygen-deficient VO$_{2-x}$ and topochemically reduced V$_2$O$_3$. **c**, Heterointerface region of cross-sectional high-resolution transmission electron microscopy (HRTEM) images for topochemically reduced V$_2$O$_3$ deposited on *c*-plane Al$_2$O$_3$ substrate. **d**, Zoom-in HRTEM image for V$_2$O$_3$ film and respective Fast Fourier transform (FFT) images. **e**, HRTEM and **f**, corresponding zoom-in images for differently oriented domains existed in topochemically reduced V$_2$O$_3$ film.

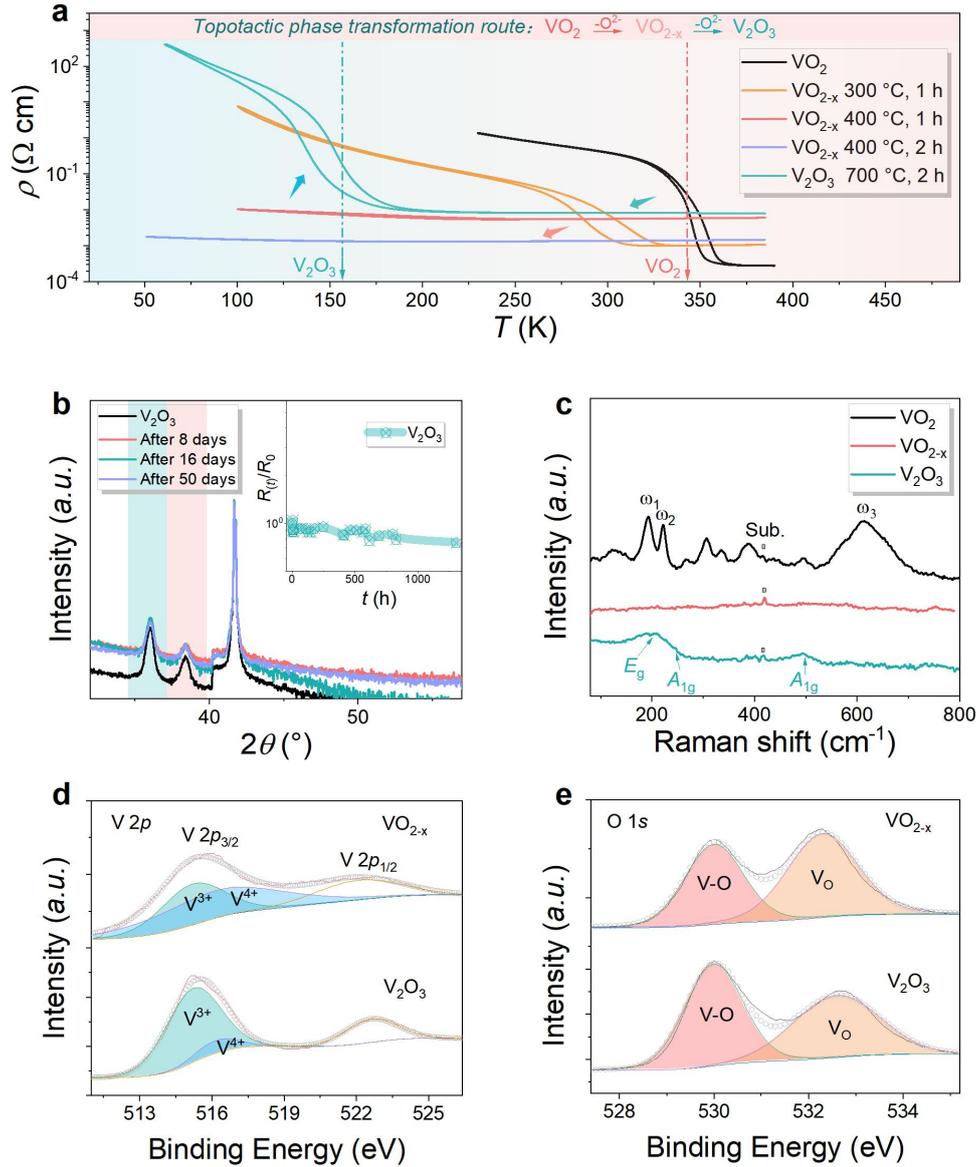

**Figure 2. Oxygen-vacancy-driven topochemical reduction in $VO_2$ system. a**, Temperature dependence of material resistivity ($\rho$-$T$) as measured for correlated $VO_2$ through oxygen vacancy ordering. **b**, Chemical stability in topochemically reduced $V_2O_3$ under ambient atmosphere. **c**, Raman spectra as compared for $VO_2$, $VO_{2-x}$ and $V_2O_3$ system through topochemical reduction. **d-e**, X-ray photoelectron spectra (XPS) compared for the core levels of **d**, vanadium and **e**, oxygen of oxygen-deficient $VO_{2-x}$ and topochemically reduced $V_2O_3$.

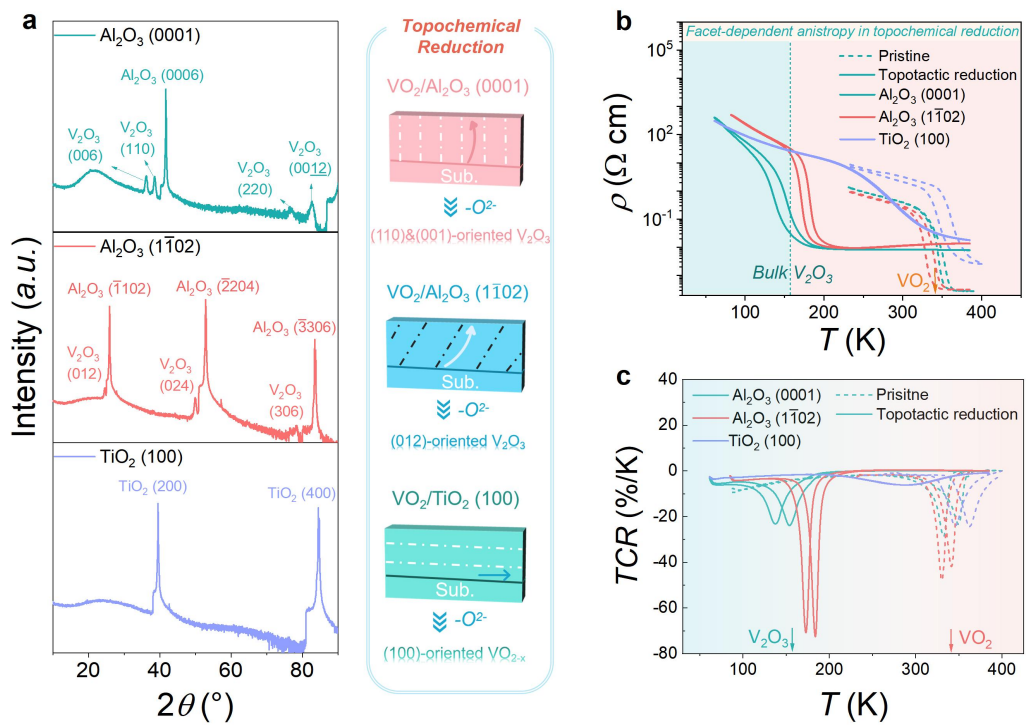

**Figure 3. Facet-dependent anisotropy in topochemical reduction of VO$_2$ system. a**, Comparing the XRD patterns for topochemical reduction of VO$_2$ deposited on differently epitaxial templates. **b**, Anisotropic transport behaviors for topochemically reduced VO$_2$. **c**, *TCR-T* tendency compared for topochemically reduced VO$_2$ deposited on different templates.

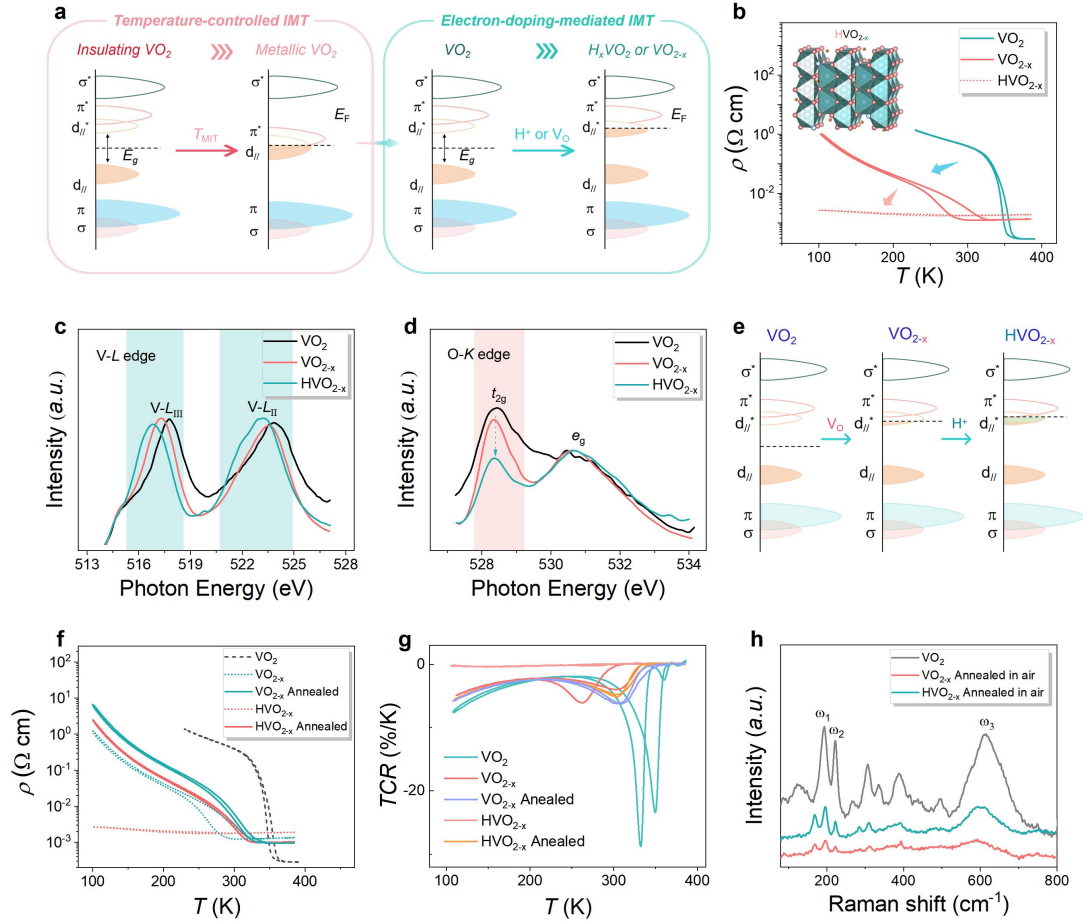

**Figure 4. Ionic interactions between oxygen vacancy and intercalated hydrogens in correlated VO₂ system. a**, Schematic of filling-controlled Mott phase modulations of VO₂ associated with oxygen vacancy and hydrogenation. **b**, $\rho$-$T$ tendency compared for oxygen-deficient VO$_{2-x}$ before and after hydrogenation. **c-d**, Soft X-ray absorption spectra (sXAS) for the **c**, V-*L* edge and **d**, O-*K* edge of oxygen-deficient VO$_{2-x}$ before and after hydrogenation. **e**, Schematic of cooperative electron doping in driving electronic phase transition of VO₂ through oxygen vacancy and hydrogenation. **f**, Comparing the oxygen ionic transport in VO₂ system through the ionic interactions. **g**, *TCR-T* tendency compared for oxygen extraction kinetics in VO₂ with and without hydrogenation. **h**, Raman spectra for annealed VO₂ with and without prior hydrogenation.

# References

(1) Lu, N.; Zhang, Z.; Wang, Y.; Li, H.-B.; Qiao, S.; Zhao, B.; He, Q.; Lu, S.; Li, C.; Wu, Y.; Zhu, M.; Lyu, X.; Chen, X.; Li, Z.; Wang, M.; Zhang, J.; Tsang, S. C.; Guo, J.; Yang, S.; Zhang, J.; Deng, K.; Zhang, D.; Ma, J.; Ren, J.; Wu, Y.; Zhu, J.; Zhou, S.; Tokura, Y.; Nan, C.-W.; Wu, J.; Yu, P. Enhanced low-temperature proton conductivity in hydrogen-intercalated brownmillerite oxide. *Nat. Energy* **2022,** *7*, 1208-1216, DOI: 10.1038/s41560-022-01166-8

(2) Li, D.; Lee, K.; Wang, B. Y.; Osada, M.; Crossley, S.; Lee, H. R.; Cui, Y.; Hikita, Y.; Hwang, H. Y. Superconductivity in an infinite-layer nickelate. *Nature* **2019,** *572*, 624-627, DOI: 10.1038/s41586-019-1496-5

(3) Lee, K.; Wang, B. Y.; Osada, M.; Goodge, B. H.; Wang, T. C.; Lee, Y.; Harvey, S.; Kim, W. J.; Yu, Y.; Murthy, C.; Raghu, S.; Kourkoutis, L. F.; Hwang, H. Y. Linear-in-temperature resistivity for optimally superconducting (Nd,Sr)NiO$_2$. *Nature* **2023,** *619*, 288-292, DOI: 10.1038/s41586-023-06129-x

(4) Nukala, P.; Ahmadi, M.; Wei, Y.; de Graaf, S.; Stylianidis, E.; Chakrabortty, T.; Matzen, S.; Zandbergen, H. W.; Björling, A.; Mannix, D.; Carbone, D.; Kooi, B.; Noheda, B. Reversible oxygen migration and phase transitions in hafnia-based ferroelectric devices. *Science* **2021,** *372*, 630-635, DOI: doi:10.1126/science.abf3789

(5) Chen, K.; Yuan, X.; Tian, Z.; Zou, M.; Yuan, Y.; Chen, Z.; Zhang, Q.; Zhang, Y.; Jin, X.; Wu, T.; Shahbazian-Yassar, R.; Liu, G. A facile approach for generating ordered oxygen vacancies in metal oxides. *Nat. Mater.* **2025,** *24*, 835-842, DOI: 10.1038/s41563-025-02171-4

(6) Jeong, J.; Aetukuri, N.; Graf, T.; Schladt, T. D.; Samant, M. G.; Parkin, S. S. P. Suppression of Metal-Insulator Transition in VO$_2$ by Electric Field-Induced Oxygen Vacancy Formation. *Science* **2013,** *339*, 1402-1405, DOI: 10.1126/science.1230512

(7) Kotiuga, M.; Zhang, Z.; Li, J.; Rodolakis, F.; Zhou, H.; Sutarto, R.; He, F.; Wang, Q.; Sun, Y.; Wang, Y.; Aghamiri, N. A.; Hancock, S. B.; Rokhinson, L. P.; Landau, D. P.; Abate, Y.; Freeland, J. W.; Comin, R.; Ramanathan, S.; Rabe, K. M. Carrier localization in perovskite nickelates from oxygen vacancies. *Proc. Natl. Acad. Sci.* **2019,** *116*, 21992-21997, DOI: doi:10.1073/pnas.1910490116

(8) Park, Y.; Sim, H.; Jo, M.; Kim, G.-Y.; Yoon, D.; Han, H.; Kim, Y.; Song, K.; Lee, D.; Choi, S.-Y.; Son, J. Directional ionic transport across the oxide interface enables low-temperature epitaxy of rutile TiO$_2$. *Nat. Commun.* **2020,** *11*, 1401, DOI: 10.1038/s41467-020-15142-x

(9) Lu, N. P.; Zhang, P. F.; Zhang, Q. H.; Qiao, R. M.; He, Q.; Li, H. B.; Wang, Y. J.; Guo, J. W.; Zhang, D.; Duan, Z.; Li, Z. L.; Wang, M.; Yang, S. Z.; Yan, M. Z.; Arenholz, E.; Zhou, S. Y.; Yang, W. L.; Gu, L.; Nan, C. W.; Wu, J.; Tokura, Y.; Yu, P. Electric-field control of tri-state phase transformation with a selective dual-ion switch. *Nature* **2017,** *546*, 124-128, DOI: 10.1038/nature22389

(10) Zhang, J.; Shen, S.; Puggioni, D.; Wang, M.; Sha, H.; Xu, X.; Lyu, Y.; Peng, H.; Xing, W.; Walters, L. N.; Liu, L.; Wang, Y.; Hou, D.; Xi, C.; Pi, L.; Ishizuka, H.; Kotani, Y.; Kimata, M.; Nojiri, H.; Nakamura, T.; Liang, T.; Yi, D.; Nan, T.; Zang, J.; Sheng, Z.; He, Q.; Zhou, S.; Nagaosa, N.; Nan, C.-W.; Tokura, Y.; Yu, R.; Rondinelli, J. M.; Yu, P. A correlated ferromagnetic polar metal by design. *Nat. Mater.* **2024,** *23*, 912-919, DOI: 10.1038/s41563-024-01856-6

(11) Tian, J.; Wu, H.; Fan, Z.; Zhang, Y.; Pennycook, S. J.; Zheng, D.; Tan, Z.; Guo, H.; Yu, P.; Lu, X.; Zhou, G.; Gao, X.; Liu, J.-M. Nanoscale Topotactic Phase Transformation in SrFeO$_x$ Epitaxial Thin Films for High-Density Resistive Switching Memory. *Adv. Mater.* **2019,** *31*, 1903679, DOI: https://doi.org/10.1002/adma.201903679

(12) Rogge, P. C.; Chandrasena, R. U.; Cammarata, A.; Green, R. J.; Shafer, P.; Lefler, B. M.; Huon, A.;

Arab, A.; Arenholz, E.; Lee, H. N.; Lee, T.-L.; Nemšák, S.; Rondinelli, J. M.; Gray, A. X.; May, S. J. Electronic structure of negative charge transfer CaFeO$_3$ across the metal-insulator transition. *Phys. Rev. Mater.* **2018,** *2*, 015002, DOI: 10.1103/PhysRevMaterials.2.015002

(13) Jeen, H.; Choi, W. S.; Biegalski, M. D.; Folkman, C. M.; Tung, I. C.; Fong, D. D.; Freeland, J. W.; Shin, D.; Ohta, H.; Chisholm, M. F.; Lee, H. N. Reversible redox reactions in an epitaxially stabilized SrCoO$_x$ oxygen sponge. *Nat. Mater.* **2013,** *12*, 1057-1063, DOI: 10.1038/nmat3736

(14) Wall, S.; Yang, S.; Vidas, L.; Chollet, M.; Glownia, J. M.; Kozina, M.; Katayama, T.; Henighan, T.; Jiang, M.; Miller, T. A.; Reis, D. A.; Boatner, L. A.; Delaire, O.; Trigo, M. Ultrafast disordering of vanadium dimers in photoexcited VO$_2$. *Science* **2018,** *362*, 572-+, DOI: 10.1126/science.aau3873

(15) Lee, D.; Chung, B.; Shi, Y.; Kim, G.-Y.; Campbell, N.; Xue, F.; Song, K.; Choi, S.-Y.; Podkaminer, J. P.; Kim, T. H.; Ryan, P. J.; Kim, J.-W.; Paudel, T. R.; Kang, J.-H.; Spinuzzi, J. W.; Tenne, D. A.; Tsymbal, E. Y.; Rzchowski, M. S.; Chen, L. Q.; Lee, J.; Eom, C. B. Isostructural metal-insulator transition in VO$_2$. *Science* **2018,** *362*, 1037-1040, DOI: doi:10.1126/science.aam9189

(16) Sangwook, L.; Hippalgaonkar, K.; Yang, F.; Hong, J.; Ko, C.; Suh, J.; Liu, K.; Wang, K.; Urban, J. J.; Zhang, X.; Dames, C.; Hartnoll, S. A.; Delaire, O.; Wu, J. Anomalously low electronic thermal conductivity in metallic vanadium dioxide. *Science* **2017,** *355*, 371-374, DOI: 10.1126/science.aag0410

(17) Zhang, H.-T.; Guo, L.; Stone, G.; Zhang, L.; Zheng, Y.-X.; Freeman, E.; Keefer, D. W.; Chaudhuri, S.; Paik, H.; Moyer, J. A.; Barth, M.; Schlom, D. G.; Badding, J. V.; Datta, S.; Gopalan, V.; Engel-Herbert, R. Imprinting of Local Metallic States into VO2 with Ultraviolet Light. *Adv. Funct. Mater.* **2016,** *26*, 6612-6618, DOI: https://doi.org/10.1002/adfm.201601890

(18) Sim, H.; Doh, K.-Y.; Park, Y.; Song, K.; Kim, G.-Y.; Son, J.; Lee, D.; Choi, S.-Y. Crystallographic Pathways to Tailoring Metal-Insulator Transition through Oxygen Transport in VO$_2$. *Small* **2024,** *20*, 2402260, DOI: https://doi.org/10.1002/smll.202402260

(19) Zhang, Z.; Zuo, F.; Wan, C. H.; Dutta, A.; Kim, J.; Rensberg, J.; Nawrodt, R.; Park, H. H.; Larrabee, T. J.; Guan, X. F.; Zhou, Y.; Prokes, S. M.; Ronning, C.; Shalaev, V. M.; Boltasseva, A.; Kats, M. A.; Ramanathan, S. Evolution of Metallicity in Vanadium Dioxide by Creation of Oxygen Vacancies. *Phys. Rev. Appl.* **2017,** *7*, DOI: 10.1103/PhysRevApplied.7.034008

(20) Hu, P.; Hu, P.; Vu, T. D.; Li, M.; Wang, S.; Ke, Y.; Zeng, X.; Mai, L.; Long, Y. Vanadium Oxide: Phase Diagrams, Structures, Synthesis, and Applications. *Chem. Rev.* **2023,** *123*, 4353-4415, DOI: 10.1021/acs.chemrev.2c00546

(21) Xue, W.; Liu, G.; Zhong, Z.; Dai, Y.; Shang, J.; Liu, Y.; Yang, H.; Yi, X.; Tan, H.; Pan, L.; Gao, S.; Ding, J.; Xu, X.-H.; Li, R.-W. A 1D Vanadium Dioxide Nanochannel Constructed via Electric-Field-Induced Ion Transport and its Superior Metal–Insulator Transition. *Adv. Mater.* **2017,** *29*, 1702162, DOI: https://doi.org/10.1002/adma.201702162

(22) Li, B.; Hu, M.; Ren, H.; Hu, C.; Li, L.; Zhang, G.; Jiang, J.; Zou, C. Atomic Origin for Hydrogenation Promoted Bulk Oxygen Vacancies Removal in Vanadium Dioxide. *J. Phys. Chem. Lett.* **2020,** *11*, 10045-10051, DOI: 10.1021/acs.jpclett.0c02773

(23) Hu, M.; Zhu, Q.; Zhao, Y.; Zhang, G.; Zou, C.; Prezhdo, O.; Jiang, J. Facile Removal of Bulk Oxygen Vacancy Defects in Metal Oxides Driven by Hydrogen-Dopant Evaporation. *J. Phys. Chem. Lett.* **2021,** *12*, 9579-9583, DOI: 10.1021/acs.jpclett.1c02687

(24) Zhang, Y.; Wang, Y. P.; Wu, Y. S.; Shu, X. Y.; Zhang, F.; Peng, H. N.; Shen, S. C.; Ogawa, N.; Zhu, J. Y.; Yu, P. Artificially controlled nanoscale chemical reduction in VO$_2$ through electron beam illumination. *Nat. Commun.* **2023,** *14*, 4012 DOI: 10.1038/s41467-023-39812-8

(25) Ruzmetov, D.; Senanayake, S. D.; Narayanamurti, V.; Ramanathan, S. Correlation between

metal-insulator transition characteristics and electronic structure changes in vanadium oxide thin films. *Phys. Rev. B* **2008,** *77*, 195442, DOI: 10.1103/PhysRevB.77.195442

(26) Park, J.; Yoon, H.; Sim, H.; Choi, S. Y.; Son, J. Accelerated Hydrogen Diffusion and Surface Exchange by Domain Boundaries in Epitaxial $VO_2$ Thin Films. *ACS Nano* **2020,** *14*, 2533-2541, DOI: 10.1021/acsnano.0c00441

(27) Kalcheim, Y.; Adda, C.; Salev, P.; Lee, M.-H.; Ghazikhanian, N.; Vargas, N. M.; del Valle, J.; Schuller, I. K. Structural Manipulation of Phase Transitions by Self-Induced Strain in Geometrically Confined Thin Films. *Adv. Funct. Mater.* **2020,** *30*, 2005939, DOI: https://doi.org/10.1002/adfm.202005939

(28) Li, L.; Wang, M.; Zhou, Y.; Zhang, Y.; Zhang, F.; Wu, Y.; Wang, Y.; Lyu, Y.; Lu, N.; Wang, G.; Peng, H.; Shen, S.; Du, Y.; Zhu, Z.; Nan, C.-W.; Yu, P. Manipulating the insulator-metal transition through tip-induced hydrogenation. *Nat. Mater.* **2022,** *21*, 1246-1251, DOI: 10.1038/s41563-022-01373-4

(29) Lee, Y.; Wei, X.; Yu, Y.; Bhatt, L.; Lee, K.; Goodge, B. H.; Harvey, S. P.; Wang, B. Y.; Muller, D. A.; Kourkoutis, L. F.; Lee, W.-S.; Raghu, S.; Hwang, H. Y. Synthesis of superconducting freestanding infinite-layer nickelate heterostructures on the millimetre scale. *Nat. Synth.* **2025,** *4*, 573-581, DOI: 10.1038/s44160-024-00714-2

(30) Zhou, X.; Yao, X.; Lu, W.; Guo, J.; Ji, J.; Lang, L.; Zhou, G.; Yao, C.; Qiao, X.; Ji, H.; Yuan, Z.; Xu, X. Manipulating the hydrogen-induced insulator-metal transition through artificial microstructure engineering. **2025**, DOI: https://doi.org/10.48550/arXiv.2505.15181

(31) Kalcheim, Y.; Butakov, N.; Vargas, N. M.; Lee, M.-H.; del Valle, J.; Trastoy, J.; Salev, P.; Schuller, J.; Schuller, I. K. Robust Coupling between Structural and Electronic Transitions in a Mott Material. *Phys. Rev. Lett.* **2019,** *122*, 057601, DOI: 10.1103/PhysRevLett.122.057601

(32) Zhou, X.; Li, H.; Jiao, Y.; Zhou, G.; Ji, H.; Jiang, Y.; Xu, X. Hydrogen‐Associated Multiple Electronic Phase Transitions for d‐Orbital Transitional Metal Oxides: Progress, Application, and Beyond. *Adv. Funct. Mater.* **2024,** *34*, 2316536, DOI: 10.1002/adfm.202316536

(33) Yoon, H.; Choi, M.; Lim, T. W.; Kwon, H.; Ihm, K.; Kim, J. K.; Choi, S. Y.; Son, J. Reversible phase modulation and hydrogen storage in multivalent $VO_2$ epitaxial thin films. *Nat. Mater.* **2016,** *15*, 1113-1119, DOI: 10.1038/nmat4692

(34) Zhou, X.; Jiao, Y.; Lu, W.; Guo, J.; Yao, X.; Ji, J.; Zhou, G.; Ji, H.; Yuan, Z.; Xu, X. Hydrogen-Associated Filling-Controlled Mottronics Within Thermodynamically Metastable Vanadium Dioxide. *Adv. Sci.* **2025,** *12*, 2414991, DOI: 10.1002/advs.202414991

(35) Chen, Y. L.; Wang, Z. W.; Chen, S.; Ren, H.; Wang, L. X.; Zhang, G. B.; Lu, Y. L.; Jiang, J.; Zou, C. W.; Luo, Y. Non-catalytic hydrogenation of $VO_2$ in acid solution. *Nat. Commun.* **2018,** *9*, 818, DOI: 10.1038/s41467-018-03292-y

(36) Zhou, X.; Shang, Y.; Gu, Z.; Jiang, G.; Ozawa, T.; Mao, W.; Fukutani, K.; Matsuzaki, H.; Jiang, Y.; Chen, N.; Chen, J. Revealing the role of high-valence elementary substitution in the hydrogen-induced Mottronic transitions of vanadium dioxide. *Appl. Phys. Lett.* **2024,** *124*, 082103, DOI: 10.1063/5.0189271